\documentclass[preprint,12pt]{elsarticle}

\usepackage{amsmath}
\usepackage{graphicx}
\usepackage{dcolumn}
\usepackage{bm}
\usepackage{amssymb}
\usepackage{graphics}
\usepackage{hyperref}
\usepackage{epsfig}
\usepackage{subfigure}
\usepackage{threeparttable}

\journal{xxx}

\begin{document}
\begin{frontmatter}


\title{Monte Carlo simulation of equilibrium and dynamic phase transition properties of an Ising bilayer}

\author[]{Yusuf Y\"{u}ksel}

\cortext[]{Corresponding author. Tel.: +902323019544; fax: +902324534188.} \ead{yusuf.yuksel@deu.edu.tr}

\address{Department of Physics, Dokuz Eyl\"{u}l University,
Tr-35160 \.{I}zmir, Turkey}


\begin{abstract}
Magnetic properties of an Ising bilayer system  defined on a honeycomb lattice with non-magnetic interlayers which interact via an indirect exchange coupling have been investigated by Monte Carlo simulation technique.
Equilibrium properties of the system exhibit ferrimagnetism with $P$-, $N$- and $Q$- type behaviors. 
Compensation phenomenon suddenly disappears with decreasing strength of indirect ferrimagnetic interlayer exchange coupling. 
Qualitative properties are in a good agreement with those obtained by effective field theory.
In order to investigate the stochastic dynamics of kinetic Ising bilayer, we have introduced two different types of dynamic magnetic fields, namely a square wave, and a sinusoidally oscillating magnetic field form.
For both field types, compensation point and critical temperature decrease with increasing amplitude and field period. Dynamic ferromagnetic region in the presence of square wave magnetic field is narrower than 
that obtained for sinusoidally oscillating magnetic field when the amplitude and the field period are the same for each type of dynamic magnetic fields.
\end{abstract}

\end{frontmatter}
\section{Introduction}\label{introduction}
Nowadays, magnetic properties of low dimensional systems in forms of graphene-like structures have attracted significant amount of interest. The reason is due to the fact that two dimensional graphene 
\cite{novoselov2,wallace,mcClure,slonczewski,geim}, and its variants \cite{cahangirov} defined on a honeycomb lattice 
exhibit a variety of interesting  electric and magnetic properties which are significantly affected by varying system size.  
After experimental realization of exactly two dimensional  monocrystalline graphitic films \cite{novoselov1} which are only a few atoms thick but stable under environmental conditions, 
theoretical and experimental research interests have been directed to the 
studies of two-dimensional
layered structures. For instance, in order to reveal the finite-temperature  properties  of  honeycomb  iridates  with  general  formula $\mathrm{A_{2}IrO_{3}}$ which exhibit strong spin-orbit coupling (SOC), 
Price and Perkins \cite{price1,price2} have performed Monte Carlo (MC) simulations based on the classical Heisenberg-Kitaev (HK) model \cite{kitaev} on a honeycomb lattice where the interactions 
between nearest neighbors are of $XX$, $YY$ or $ZZ$ type.
Very recently, it has been shown that transition metal trihalides ($\mathrm{MX_{3}}$) defined on a two dimensional honeycomb lattice may exhibit magnetic order below a finite critical temperature \cite{sarikurt,ersan}. 

Importance of honeycomb lattice not only originates as a consequence of experimental research on graphene, but resides also on the theoretical grounds. Namely, it offers reduced mathematical complexity, and there are also
some exact results regarding the magnetic properties for this structure \cite{horiguchi,urumov1,urumov2,urumov3,urumov4,zhen}. 
From the experimental point of view, single layer, double layer and few (3 to 10) layer honeycomb structures are classified as three different types of 2D crystals, and thin film limit is reached for thicker systems \cite{novoselov1}.   
In this regard, investigation of magnetic properties of graphene-like multilayers gained particular attention, 
and a wide variety of such systems have been successfully modeled within the framework of Ising model and its variants \cite{masrour,mhirech,jiang,santos,kaneyoshi1}. For instance, using the effective field
theory (EFT) formalism, Jiang and coworkers \cite{jiang} investigated the magnetic properties such as magnetization and the magnetic susceptibility of a nano-graphene bilayer. 
For a  trilayer Ising nanostructure, EFT calculations have been performed and from the thermal variations of the total magnetization, six distinct compensation types have been reported by Santos and S\'{a} Barreto \cite{santos}. 
In a recent paper, Kaneyoshi \cite{kaneyoshi1} investigated the magnetic behavior of an Ising bilayer with non-magnetic inter-layers. Based on EFT method, some  characteristic features of ferrimagnetism have also been reported in this study.
In that work, a realistic case has also been considered by assuming a distance-dependent indirect exchange interaction between the two magnetic layers.

On the other hand, after experimental realization of dynamic phase transitions \cite{acharyya,rikvold} in uniaxial cobalt films \cite{berger}, stochastic dynamics of 
kinetic systems gained renewed interest \cite{yuksel,vatansever}. In such systems, a dynamic phase 
transition between dynamically ordered and disordered phases takes place which is characterized by a dynamic symmetry breaking. Depending on the two competing time scales, namely, the period of the externally applied oscillating magnetic field
and relaxation time of the system, kinetic Ising model may exhibit dynamic ferromagnetic (FM) or dynamic paramagnetic (PM) character. Winner of the competition of the above mentioned time scales is determined by another complicated competition
between the field amplitude, field period, temperature, and exchange coupling.

The effective field theory \cite{kaneyoshi2} partially takes into account the spin fluctuations, and it is superior to conventional mean field theory \cite{strecka} 
where the spin-spin correlations are completely ruled out. Despite its mathematical simplicity, mean field predictions
are only valid for the systems with dimensionality $d\geq4$. 
In a recent work, we have shown that EFT and MC results qualitatively agree well with each other for a particular ternary spin system \cite{yuksel2}. 
In this regard, EFT method promises reasonable results with less computational cost. 

The aim of the present paper is two fold: First, a direct comparison of MC results obtained within the present work with the available EFT results of 
Ref. \cite{kaneyoshi1} will be presented for the Ising bilayer system. As will be shown in the following discussions, qualitatively plausible agreement exists between EFT and MC results. Second, we will present some results regarding the 
stochastic dynamics and compensation behavior of the kinetic Ising bilayer in the presence of two different forms of the oscillating magnetic field, namely a square wave form and a sinusoidal wave form. 
The rest of the paper can be outlined as follows: In Section \ref{formulation}, we will present the formulation and simulation details of our model. Section \ref{results} contains numerical results and related discussions. 
Finally, Section \ref{conclusion} is devoted to our conclusions.

\section{Model and Formulation}\label{formulation}
\begin{figure}[!h]
\center
\subfigure[\hspace{0cm}] {\includegraphics[width=4.5cm]{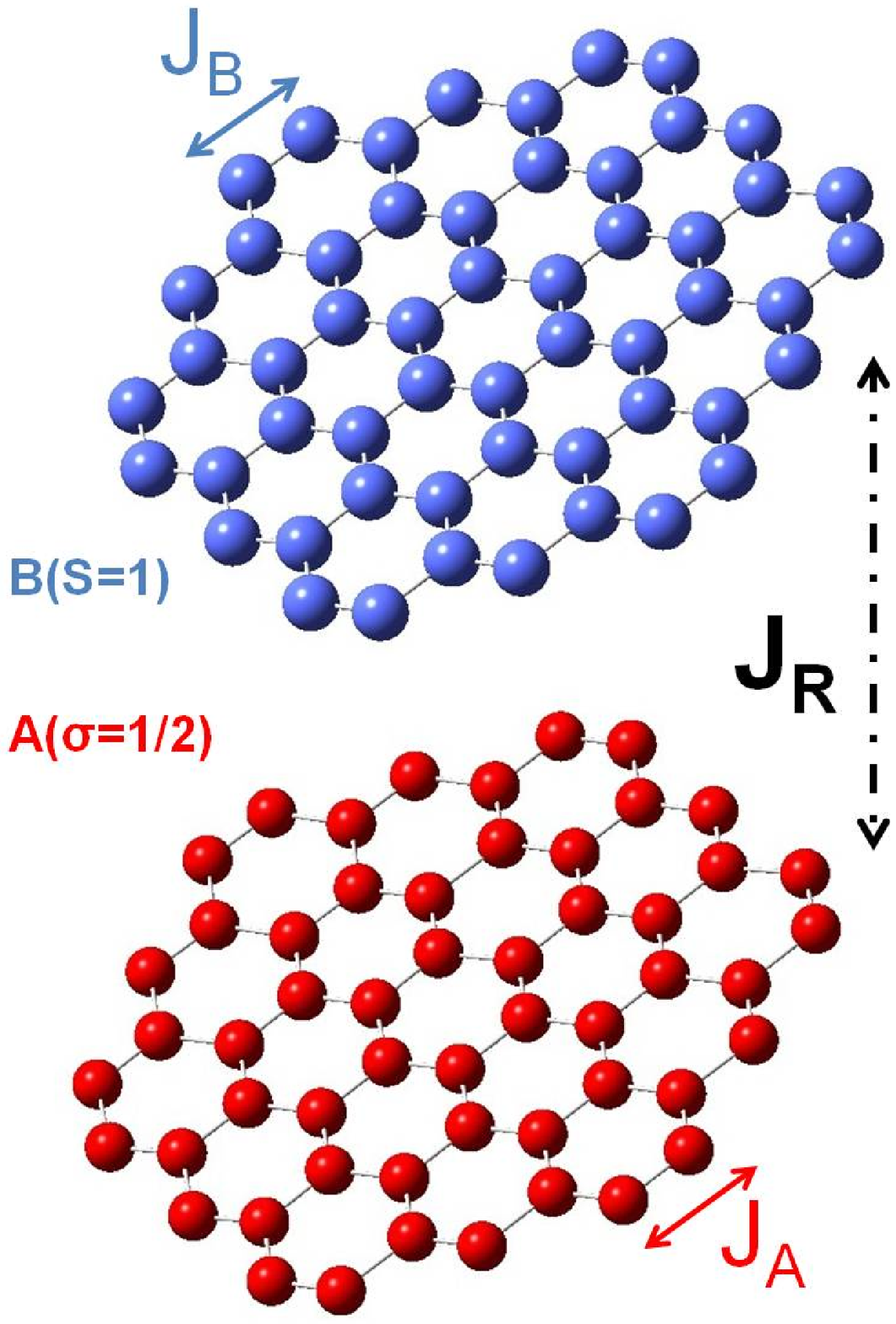}}
\hspace{0.5cm}
\subfigure[\hspace{0cm}] {\includegraphics[width=6.0cm]{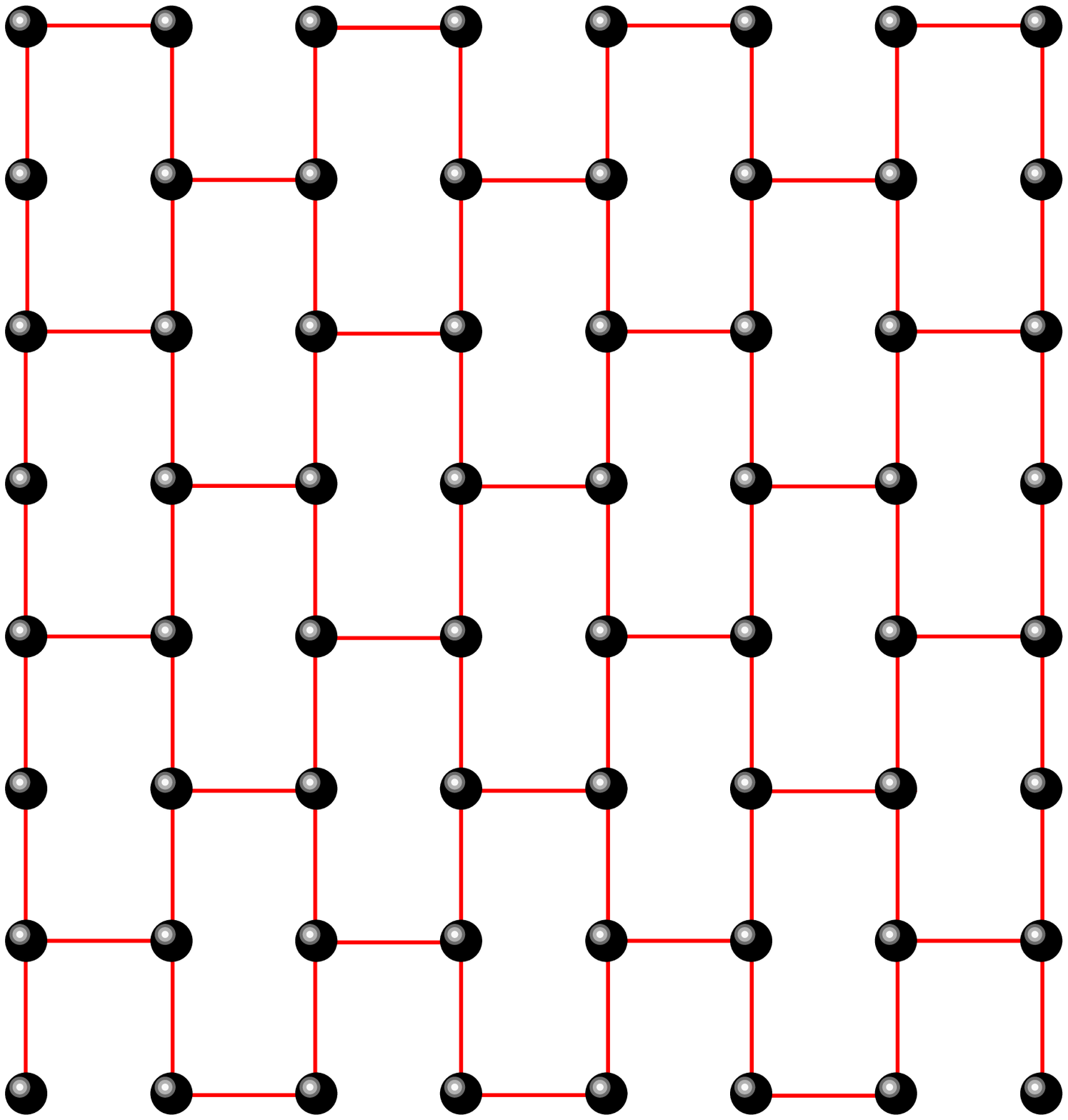}}\\
\caption{(a) Schematic representation of the simulated magnetic bilayer. Sublattice $A$ $(B)$ is occupied by $\sigma=\pm1/2$ $(S=\pm1,0)$ spins. (b) Equivalent of honeycomb lattice in the brick lattice representation. 
Each pseudo spin has three nearest neighbors, and is located on the nodes of a $L\times L$ square lattice.} \label{fig1}
\end{figure}
Our bilayer model consists of successive stacking of 2D honeycomb monolayers forming a 3D graphite structure (Fig \ref{fig1}a). 
The bottom layer, i.e. the sublattice $A$ consists of Ising spins with $\sigma_{i}=\pm\frac{1}{2}$ whereas the topmost layer (sublattice $B$)
consists of tightly packed magnetic atoms with a pseudo spin variable $S_{i}=\pm1,0$. The number of nonmagnetic layers between the sublattices $A$ and $B$ is denoted by $n$. 
The intra-layer exchange couplings are respectively denoted by $J_{A}$ $(>0)$ and $J_{B}$ $(>0)$ whereas the interlayer exchange coupling is represented by $J_{R}$ $(>0)$. 
This selection of interaction constants allows us to study the ferrimagnetic behavior of the model. 
We consider an indirect exchange coupling between the layers $A$ and $B$. 
Hence, following the same notation with Ref. \cite{kaneyoshi1}, we assume
\begin{equation}\label{eq1}
J_{R}=J\exp[-\lambda(n+1)]/(n+1)^{\delta}, 
\end{equation}
where the parameter $\lambda$ is related to the disorder and $\delta$ is related to the dimensionality of the system, and $n$ is the number of nonmagnetic layers between the sublattices $A$ and $B$, 
(please see Ref. \cite{kaneyoshi1} for details).
The Hamiltonian of the model represented by Fig. \ref{fig1} is given by 
\begin{equation}\label{eq2}
\mathcal{H}=-J_{A}\sum_{<ij>}\sigma_{i}\sigma_{j}-J_{B}\sum_{<kl>}S_{k}S_{l}+J_{R}\sum_{<ik>}\sigma_{i}S_{k}-D_{B}\sum_{k}(S_{k})^{2}, 
\end{equation}
where the spin-spin coupling terms in the first three sums are taken over only the nearest-neighbor spin pairs whereas the 
last summation is carried out over all the lattice sites of sub-lattice $B$ with $D_{B}$ being the single ion anisotropy parameter of spin-1. 

In order to implement the MC simulation procedure for the present system, each pseudo spin variable $\sigma_{i}$ and $S_{k}$ is assigned on the lattice sites of a brick lattice \cite{kim,morita} with lateral dimension $L$
which is topologically equivalent of the honeycomb lattice (Fig. \ref{fig1}b). Periodic boundary conditions have been imposed in both lateral and vertical directions. During the simulations, we have monitored the 
quantities of interest over $250000$ Monte Carlo steps per lattice site for equilibrium system, after discarding the first $50000$ steps. On the other hand, for the calculation of kinetic properties, we have 
obtained time series of magnetization over $2000$ cycles of external magnetic field, and allowed the system to relax during the first $1000$ periods.  

In the equilibrium case, the thermal average of sub-lattice ($M_{A}$ and $M_{B}$) and total $(M_{T})$ magnetizations have been calculated according to
\begin{equation}\label{eq3}
M_{\alpha}=\left\langle \sum_{t}m_{\alpha}(t) \right\rangle, \quad \alpha=A,B,T
\end{equation}
where $m_{\alpha}(t)$ is the time series of corresponding sub-lattice (or total) magnetization per spin. 
Then the definition of magnetic susceptibility and the alternative description of the total magnetization can also be given by
\begin{equation}\label{eq5}
\chi=\frac{N_{T}}{k_{B}T}\left[\left\langle \sum_{t}(m_{T}(t))^{2}\right\rangle-\left\langle \sum_{t}m_{T}(t) \right\rangle^{2}\right],
\end{equation}
\begin{equation}\label{eq4}
M_{T}=[M_{A}+M_{B}]/2.0,
\end{equation}
where $N_{T}$ is the total number of lattice sites.
Some of the simulation parameters have been fixed as $J_{A}=1.0J$, $J_{B}=0.5J$. For simplicity, we also set $k_{B}=1$.
\section{Results and Discussion}\label{results}
In section \ref{sub1}, we will present the magnetic properties of Ising bilayer in the absence of magnetic field. However, section \ref{sub2} is devoted for the discussions regarding the nonequilibrium stochastic behavior of the 
system in the presence of time dependent oscillating magnetic field.
\subsection{Equilibrium properties}\label{sub1}
\begin{figure}[!h]
\center
\subfigure[\hspace{0cm}] {\includegraphics[width=6.5cm]{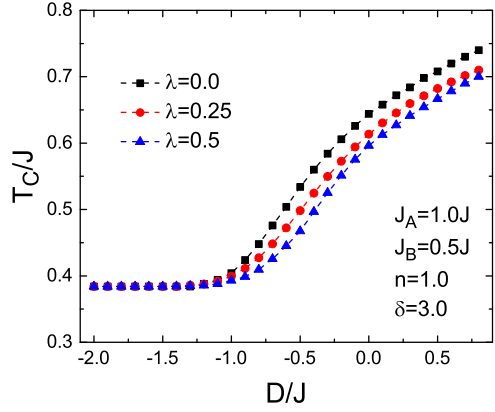}}\\
\subfigure[\hspace{0cm}] {\includegraphics[width=7.0cm]{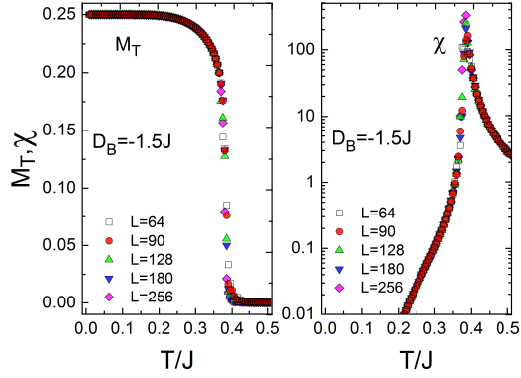}}\\
\caption{(a) Phase diagram of the Ising bilayer with $L=128$ plotted in a  $(D_{B}/J \ \mathrm{vs} \ T_{c}/J )$ plane for three different values of $\lambda$. 
(b) Magnetic properties such as the total magnetization $M_{T}$ and magnetic susceptibility $\chi$ for $D_{B}=-1.5$ with $n=1$, $\lambda=0$ and $\delta=3.0$. Different symbols correspond to different lattice size $L$.}\label{fig2}
\end{figure}
We start our investigation by examining the phase diagram of the present model in a $(D_{B}/J \ \mathrm{vs} \ T_{c}/J )$ plane for three values of disorder parameter $\lambda$ where the numerical value of the transition temperature
has been estimated from the peak point of susceptibility curves. Here, we consider one monolayer of nonmagnetic sites.
According to Eq. (\ref{eq1}), antiferromagnetic interface exchange coupling $J_{R}$ exponentially approaches to zero with increasing $\lambda$. Hence, for large values of this parameter, we have $J_{R}\rightarrow0$, and in this limit,
the two sublattices $A$ and $B$ become magnetically independent of each other. For moderate values such as $\lambda\leq0.5$, ferrimagnetic character is adopted in the system, and both sublattices undergo a phase transition at the same critical temperature.
For $\lambda=0.0$, $J_{R}$ approaches its maximum value, and for positive $D_{B}/J$, critical temperature becomes reduced with increasing $\lambda$. On the other hand, for large negative values of $D_{B}/J$, only $S_{i}=0.0$ state is allowed 
in sublattice $B$. Therefore, if we define a threshold value $D_{B}^{*}/J$ for single ion anisotropy parameter then the sublattice $B$  becomes nonmagnetic for $D_{B}/J<D_{B}^{*}/J$. In this case, the horizontal line in the phase diagram 
is the sole contribution of sublattice $A$ to the transition temperature. For spin-1 Blume-Capel model, MC calculations predict a tricritical point at $D_{t}/J=-1.446$ for the same phase diagram  \cite{booth}  
whereas EFT result is $D_{t}/J=-1.41$ \cite{kaneyoshi2,tucker}. We note that, the selection of exchange coupling parameters, namely, $J_{A}=1.0J$ and $J_{B}=0.5J$ helps us to omit the first order phase transitions in the present system. 
This can be seen from Fig. \ref{fig2}b, where we plot the magnetization and magnetic susceptibility as a function of temperature for several lattice sizes ranging from $L=64$ to $L=256$. As shown in this figure,
the magnetization exhibits a continuous phase transition in the vicinity of critical temperature and magnetic susceptibility curves exhibit a size dependent positive divergence around $T_{c}$.
All these observations clearly demonstrate that the phase transition is always of second order for the whole range of $D_{B}/J$ values. 
Besides, the ground state magnetization saturates at $M_{T}=0.25$, since the magnetization of sublattice $B$ is zero for $D_{B}=-1.5J$. 
As a final note regarding this figure, we should point out that a qualitatively similar phase diagram
has been obtained in Ref. \cite{kaneyoshi1} where the author used EFT. This fact again shows that the models solved by EFT method exhibit the same topology as those obtained from
the Monte Carlo (MC) simulation.

\begin{figure}[!h]
\center
\includegraphics[width=6.5cm]{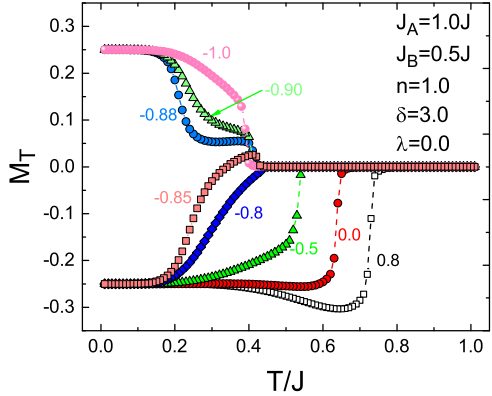}
\caption{Total magnetization $M_{T}$ as a function of $D_{B}/J$ for fixed system parameters which are shown the figure. The system size has been fixed as $L=128$. } \label{fig3}
\end{figure}
Next in Fig. \ref{fig3}, we present some ferrimagnetic properties of the system where the total magnetization $M_{T}$ has been plotted as a function of temperature for some selected values of $D_{B}/J$. The other system parameters have been fixed as
displayed in the figure. In a recent work \cite{jiang}, six different compensation types \cite{neel,strecka2} have been observed for an Ising trilayer system. On the other hand, Ref. \cite{kaneyoshi1} reports that the total magnetization of Ising bilayer with indirect 
interlayer exchange exhibits $P$-, $N$- and $Q$- type behaviors which have also been observed in our calculations. Moreover, the unclassified curve corresponding to $d=-0.85$ of Ref. \cite{kaneyoshi1} is identical to 
the curve corresponding to $D_{B}/J=-0.88$ in Fig. \ref{fig3} of the present study \cite{footnote}. This observation again supports the consistency of the results obtained by EFT and MC methods.

\begin{figure}[!h]
\center
\subfigure[\hspace{0cm}] {\includegraphics[width=6.5cm]{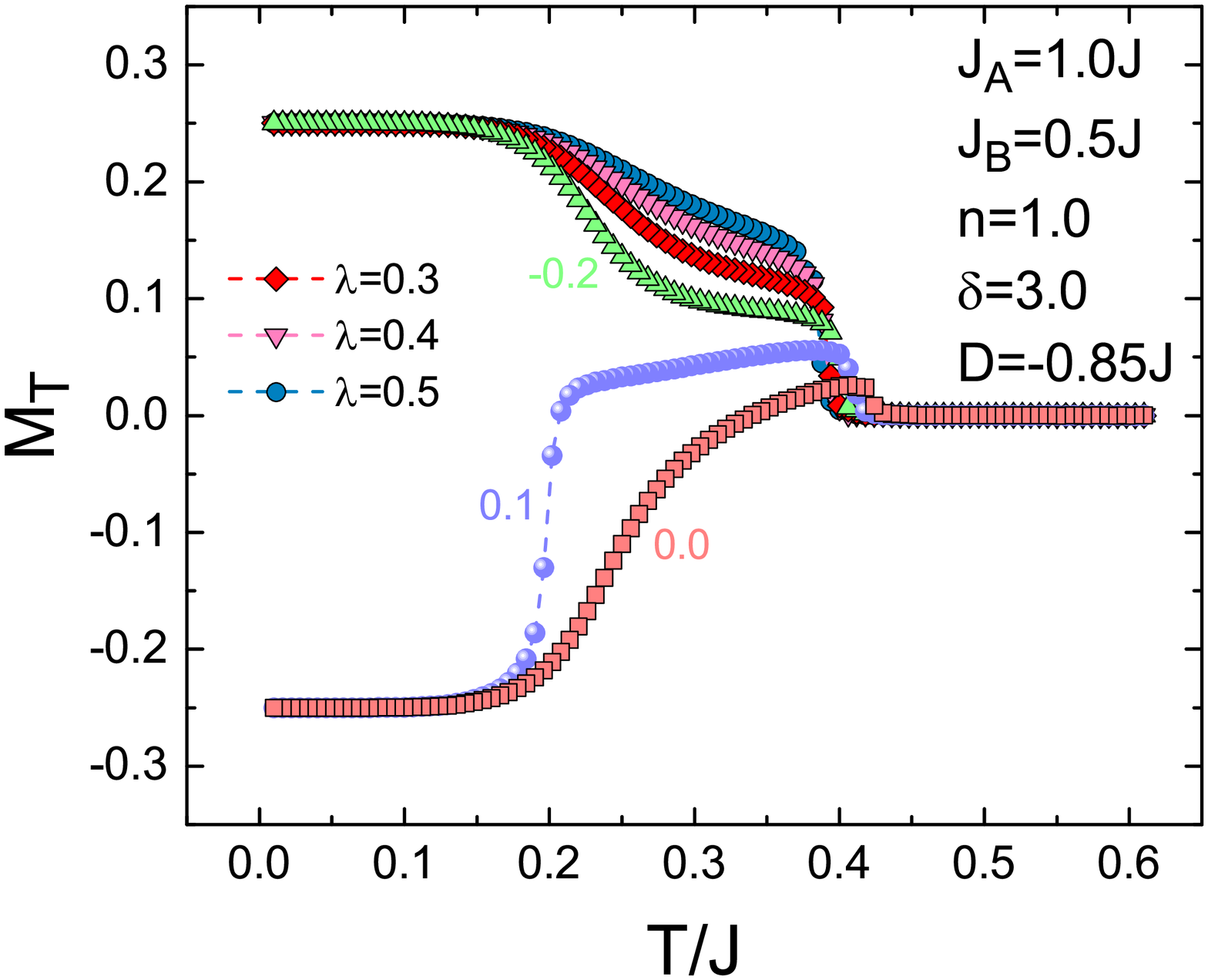}}
\subfigure[\hspace{0cm}] {\includegraphics[width=6.5cm]{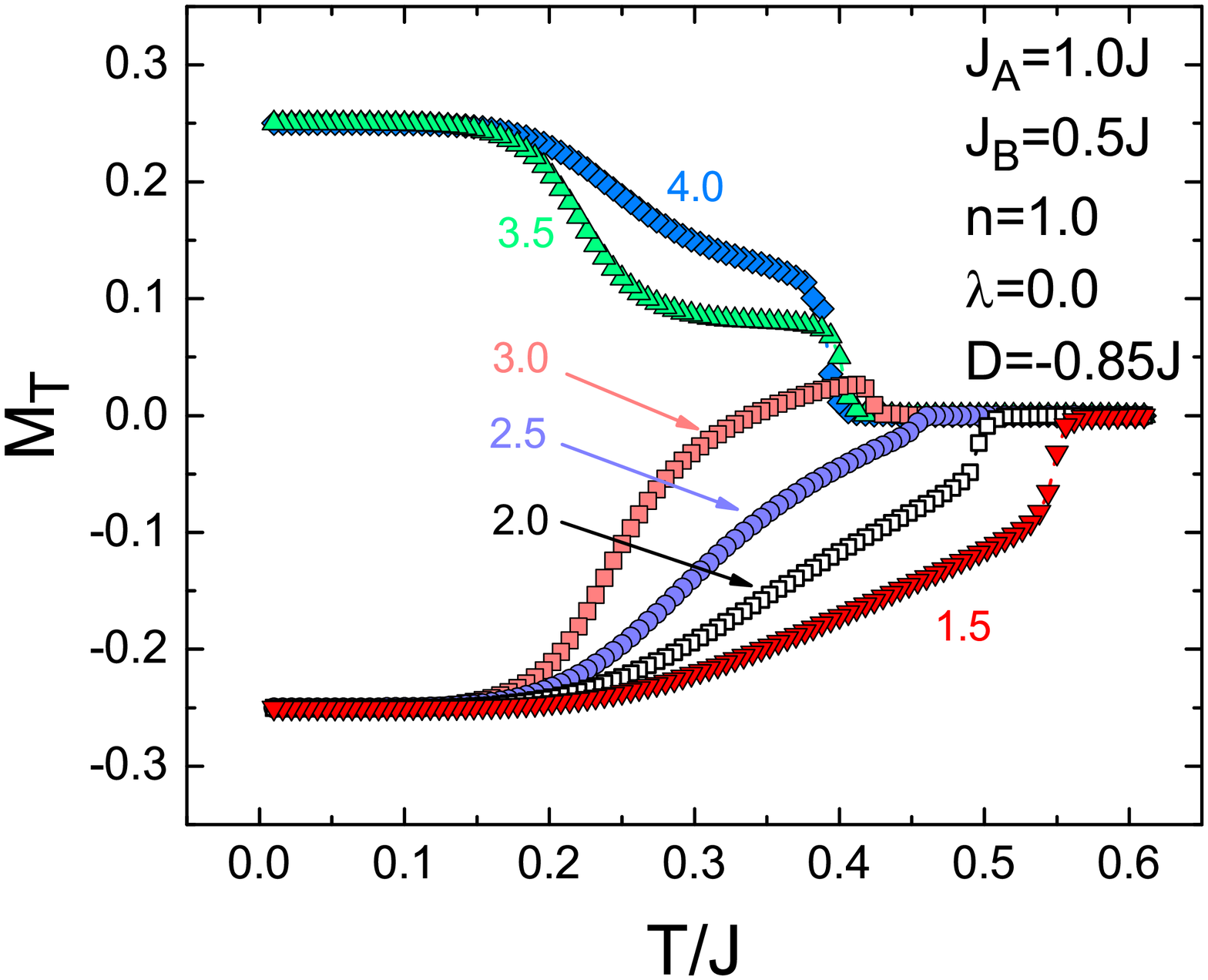}}\\
\subfigure[\hspace{0cm}] {\includegraphics[width=6.5cm]{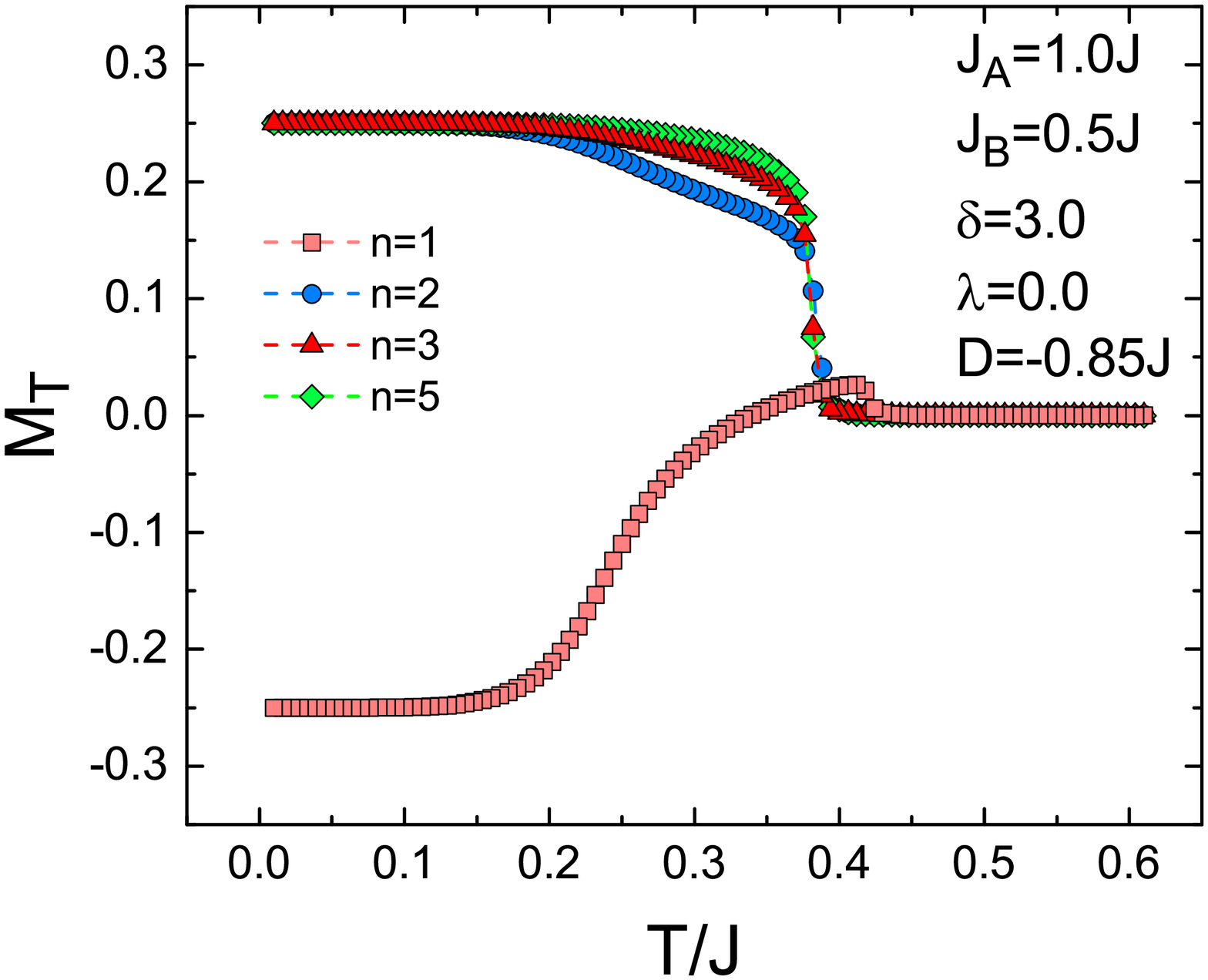}}\\
\caption{Influence of (a) $\lambda$, (b) $\delta$, and (c) $n$ on the compensation behavior of the total magnetization of the Ising bilayer with $L=128$.} \label{fig4}
\end{figure}
As shown in Fig. \ref{fig3}, a compensation behavior may originate in the system for a narrow range of $D_{B}/J$ values. 
Compensation temperature is peculiar to the systems exhibiting ferrimagnetism at which the sublattice magnetizations cancel each other below the transition temperature. 
The influence of varying $\lambda$, $\delta $ and $n$ on the magnetisation profile has been depicted in Fig. \ref{fig4}. As shown in this figure, $N$- type magnetization curve evolves towards the $Q$- type behavior with increasing 
$\lambda$, $\delta$, and $n$. This is an expected result, since $J_{R}$ rapidly decays towards zero with increasing values of these parameters. Therefore, ferrimagnetism is destructed, and we obtain two independent ferromagnetic
layers.

\subsection{Kinetic properties}\label{sub2}
\begin{figure}[!h]
\center
\subfigure[\hspace{0cm}] {\includegraphics[width=6.5cm]{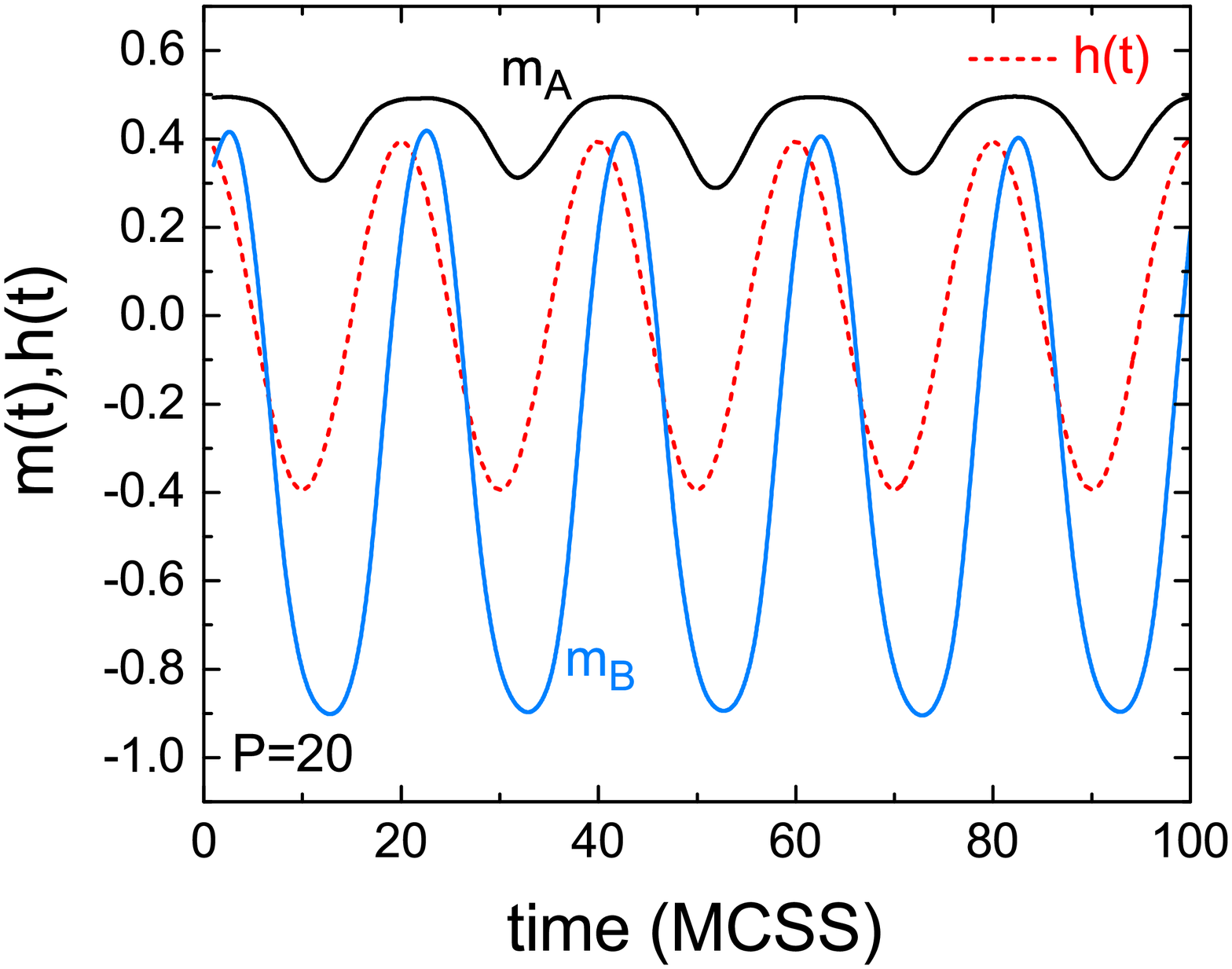}}
\subfigure[\hspace{0cm}] {\includegraphics[width=6.5cm]{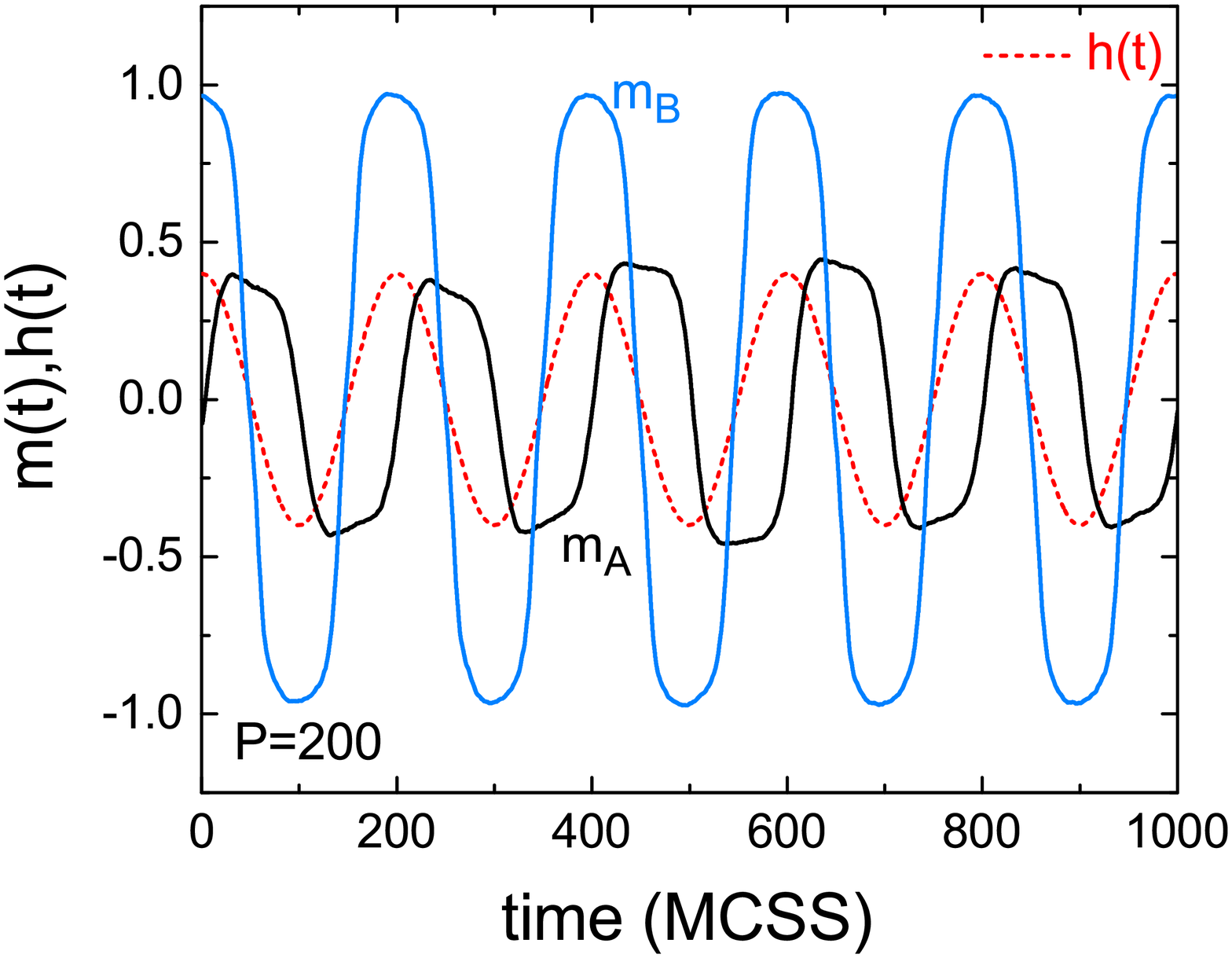}}\\
\subfigure[\hspace{0cm}] {\includegraphics[width=6.5cm]{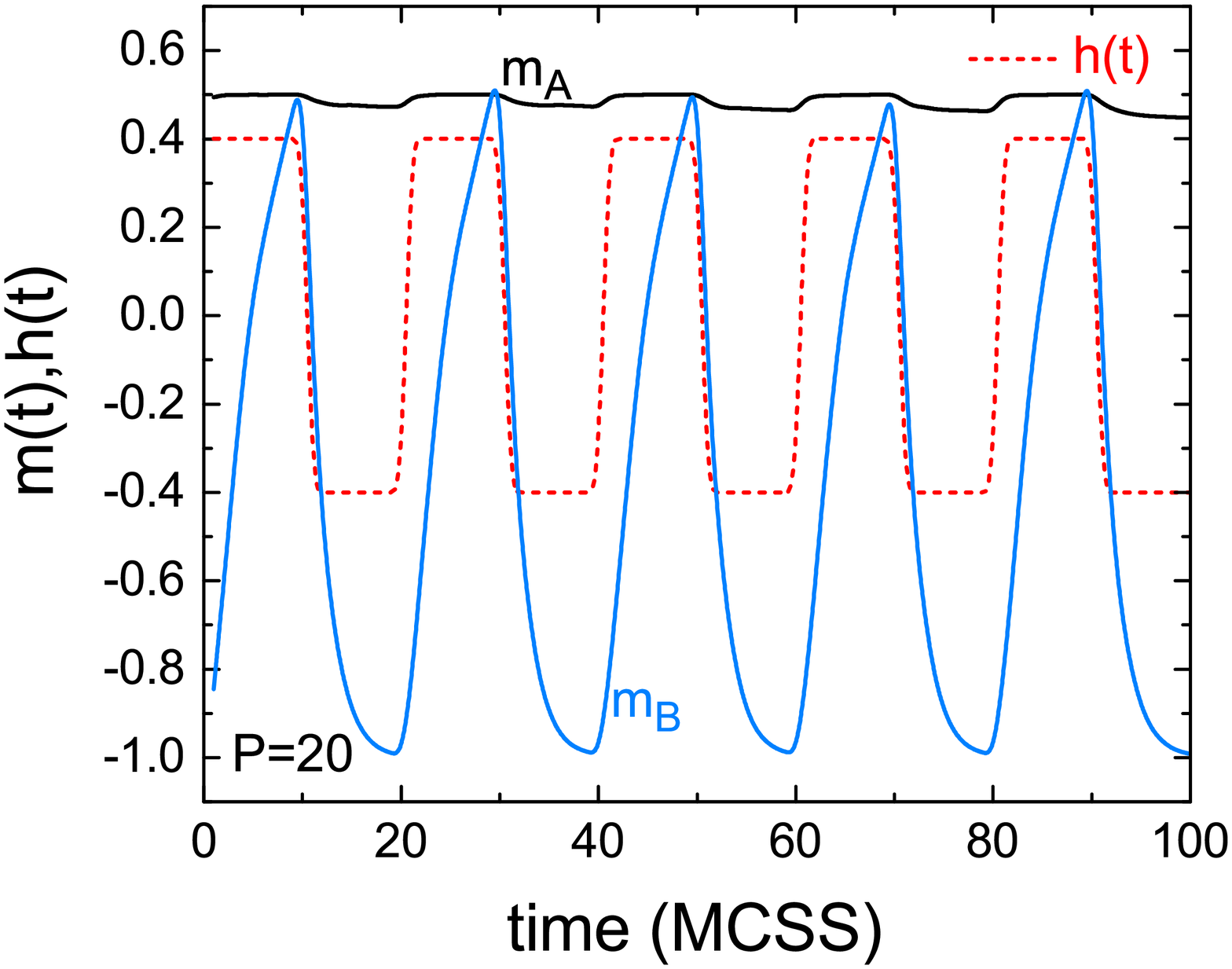}}
\subfigure[\hspace{0cm}] {\includegraphics[width=6.5cm]{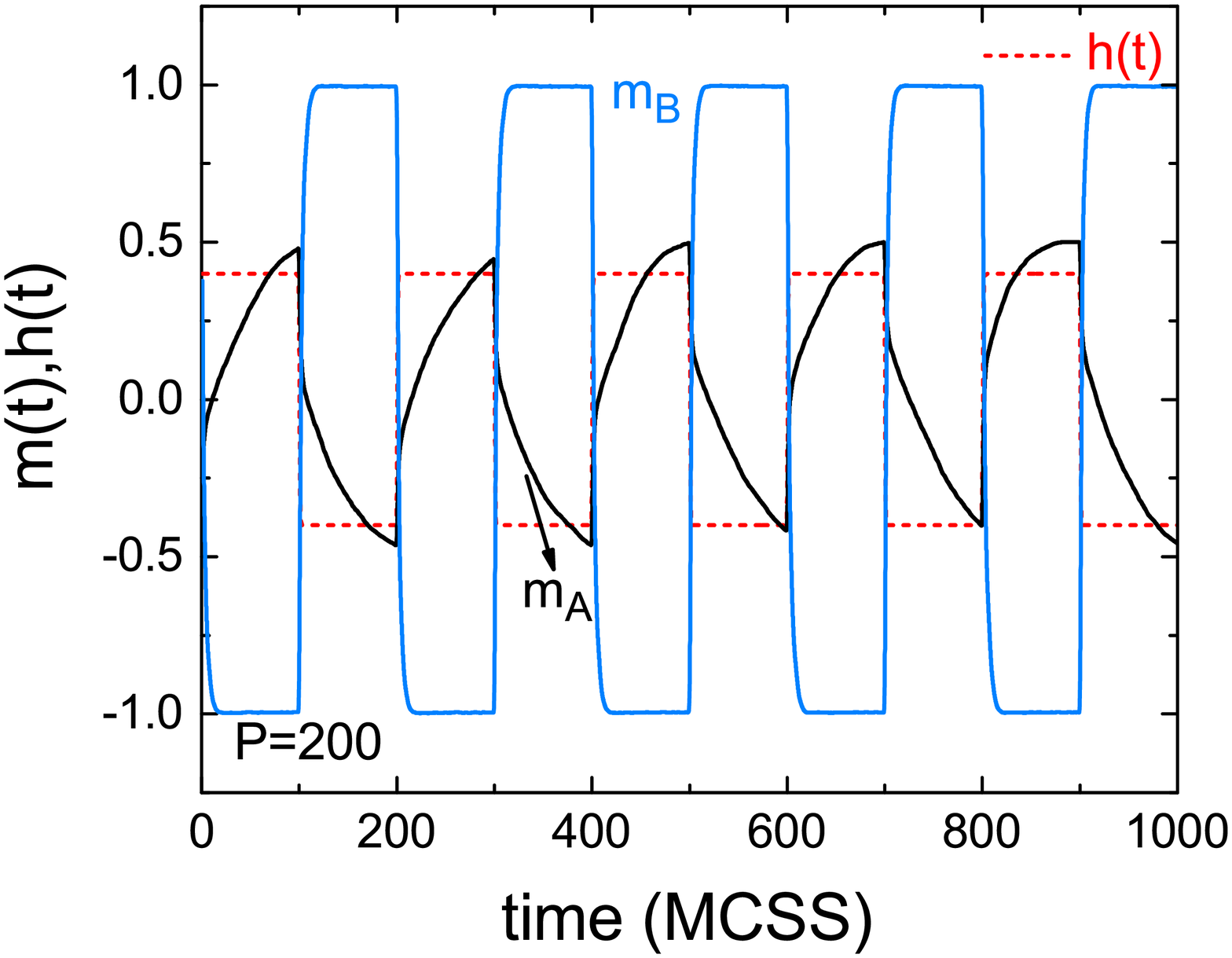}}\\
\caption{Time series of magnetizations $m_{A}$, $m_{B}$ and magnetic field $h(t)$ for the system size $L=128$. The time evolution of magnetic field is either in sinusoidal form ((a),(b)) or in square wave form ((c),(d)). 
The leftmost plots have been obtained for $P=20$ whereas the rightmost curves correspond to high period case $P=200$. The magnetic field amplitude has been fixed as $h_{0}=0.4J$.} \label{fig5}
\end{figure}
Up to now, we have considered the ferrimagnetic properties of Ising bilayer in the absence of magnetic field. 
From now on, we will discuss the variation of magnetic properties of the system in the presence of time dependent oscillating magnetic field for the following set of system parameters: 
$J_{A}=1.0J$, $J_{B}=0.5J$, $D_{B}=-0.85J$, $n=1.0$, $\delta=3.0$, and $\lambda=0.0$. This set of parameters not only allows us to avoid the first order phase transitions, but also provides information
about how the compensation behavior varies in the presence of oscillating magnetic field. For this aim we consider two distinct types of magnetic field: (i) sinusoidal wave, (ii) square wave. 
In this case, the Hamiltonian equation can be written as
\begin{equation}\label{eq7}
\mathcal{H}=\mathcal{H}_{0}+h(t)(\sum_{i}\sigma_{i}+\sum_{k}S_{k}),
\end{equation}
where $\mathcal{H}_{0}$ is the Hamiltonian equation in the absence of dynamic magnetic field, and the second and the third summations correspond to dynamic Zeeman energy terms. 
As we have underlined in the preceding sections, the system can exhibit a field induced dynamic phase transition between ordered and disordered
phases. Such a situation is shown in Fig. \ref{fig5} where we respectively select the field amplitude and the temperature as $h_{0}/J=0.4$, and $T=0.8T_{c}$. Here $T_{c}$ denotes the critical temperature in the absence of any magnetic field. 
Oscillation period of the magnetic field is denoted by $P$. In Fig. \ref{fig5}, the top and bottom panels respectively correspond to sinusoidal and square wave forms of the oscillating magnetic field. In the high frequency regime 
(i.e. the left panels) the sublattice magnetizations $m_{A}$ and $m_{B}$ oscillate around a nonzero value. This corresponds to the dynamically ordered phase. On the other hand, in the low frequency regime,
the sublattice magnetizations $m_{A}$ and $m_{B}$ can follow the external perturbation with a small phase lag, and the time average of the magnetization is very close to zero where the system is in the dynamically disordered phase. 
In this process, it is possible to trigger a field induced dynamic phase transition by properly adjusting the field period $P$.

\begin{figure}[!h]
\center
\subfigure[\hspace{0cm}] {\includegraphics[width=6.5cm]{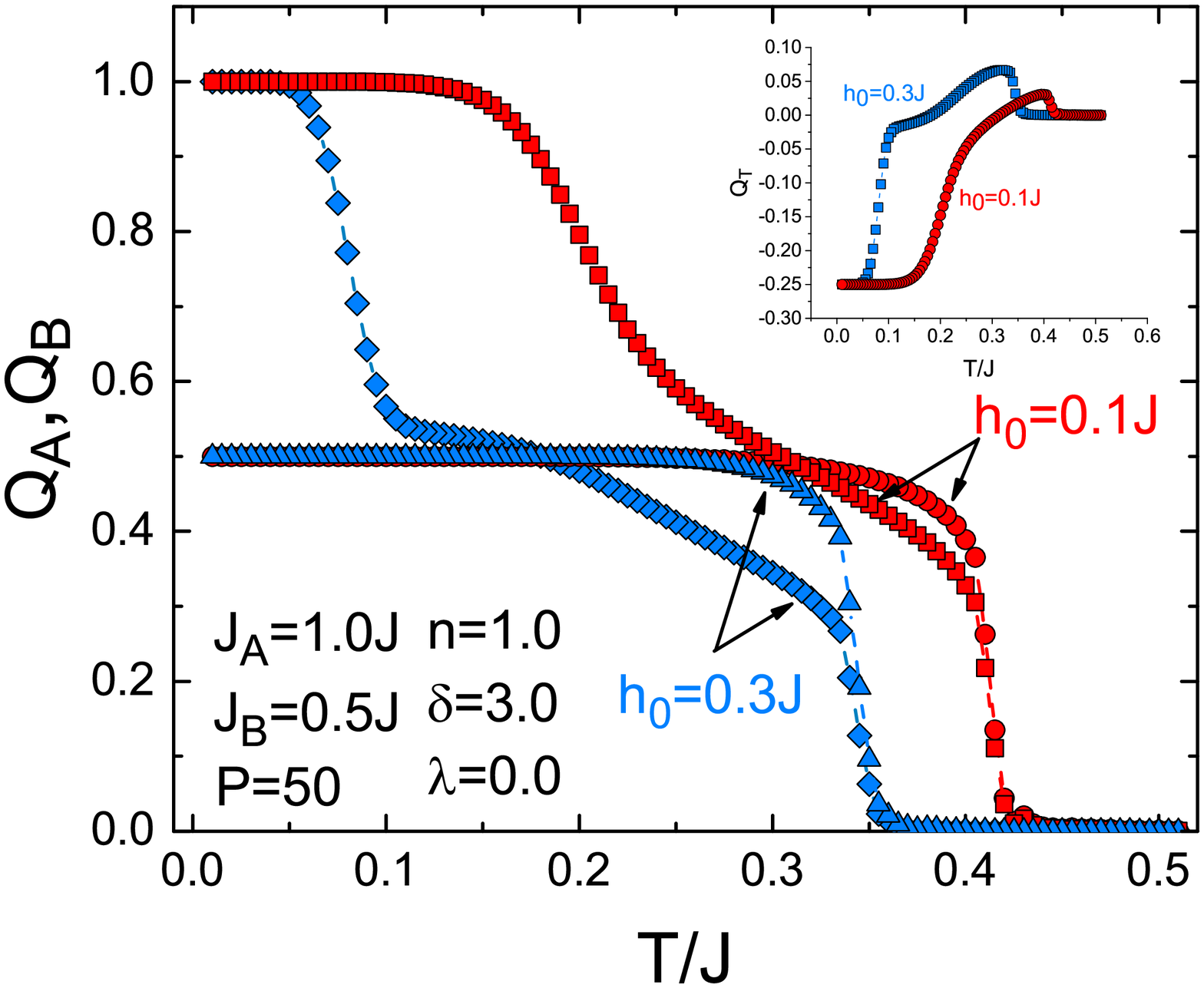}}
\subfigure[\hspace{0cm}] {\includegraphics[width=6.5cm]{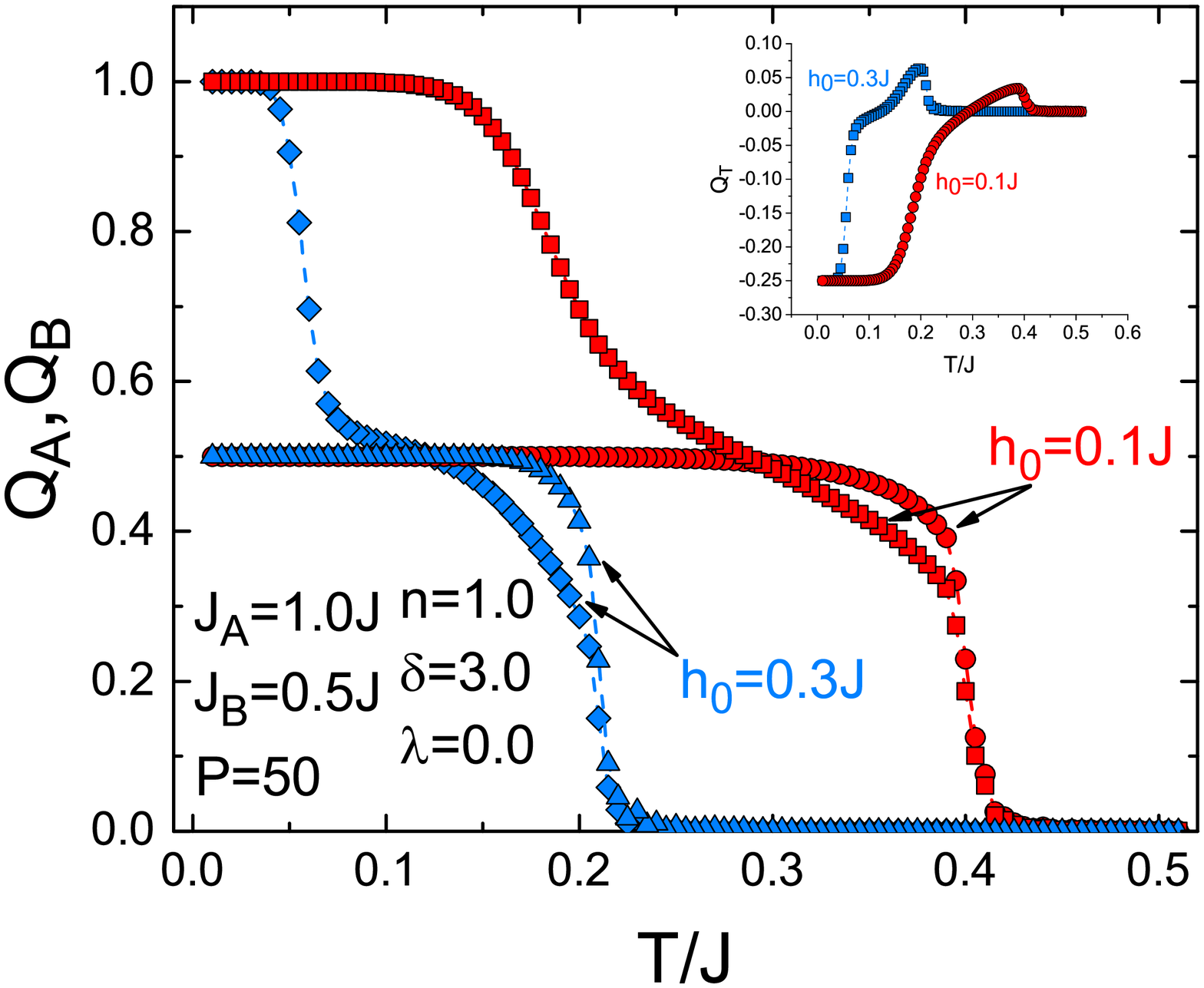}}\\
\caption{Variation of dynamic order parameters $Q_{A}$ and $Q_{B}$ as functions of temperature for $L=128$. The magnetic field $h(t)$ varies in (a) sinusoidal (b) square wave form with time. 
System parameters accompany each figure. In the inset, the total dynamic order parameter $Q_{T}$ has been depicted.} \label{fig6}
\end{figure}
Compensation behavior in the presence of dynamic magnetic fields can be examined by calculating the thermal average of dynamic order parameters corresponding to sublattices, as well as the total magnetization. These magnetic properties are defined as
the time averaged magnetizations over the successive cycles of the oscillating field \cite{tome},
\begin{equation}\label{eq8}
Q_{\alpha}=\frac{1}{NP}\oint m_{\alpha}(t)dt, \ \alpha=A,B \ \mathrm{or} \ T  
\end{equation}
where $P$ is the field period, and $N$ denotes the number of magnetic field cycles. In Fig. \ref{fig6}, in order to compare the stochastic behavior of the system in the presence of sinusoidal and square wave magnetic field,  
we have depicted the thermal variation of sublattice magnetizations $Q_{A}$ and $Q_{B}$ as functions of the temperature. It can be seen from this figure that transition temperature, as well as the compensation point $T_{comp}$
reduces with increasing magnetic field amplitude $h_{0}$. Moreover, in the presence of square wave magnetic field, numerical values of $T_{c}$ and $T_{comp}$ are clearly lower than those obtained for the sinusoidally oscillating 
magnetic fields. The insets in Fig. \ref{fig6} show the thermal variation of total magnetization when the field amplitude is varied. For both forms of the magnetic field, $Q_{T}$ exhibits $N$- type behavior. Therefore, we can conclude that 
although the compensation temperature is reduced with increasing $h_{0}$, $Q_{T}$ maintains its Ne\'{e}l classification scheme. 

\begin{figure}[!h]
\center
\subfigure[\hspace{0cm}] {\includegraphics[width=6.5cm]{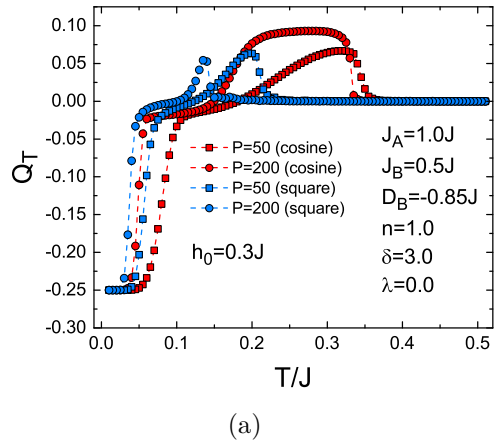}}
\caption{Variation of dynamic order parameter $Q_{T}$ as a function of temperature for $L=128$. The magnetic field $h(t)$  varies with time either in cosine or in square wave form . 
System parameters accompany each figure.} \label{fig7}
\end{figure}
Finally, let us conclude our investigation for the Ising bilayer system by examining the variation of compensation phenomenon as a function of varying field period $P$.
In figure \ref{fig7}, termal variation of $Q_{T}$ has been depicted for both sinusoidal and square wave forms of magnetic field. Here, the field amplitude has been fixed as $h_{0}=0.4J$, and we consider two different values of field period $P$.
Either for square and sinusoidal wave forms of magnetic field, the order parameter $Q_{T}$ maintains its $N$- type profile for high and low frequency perturbations.  
Our simulation results also show that increasing magnetic field period causes a decline in critical and compensation temperature values.  However, in Ref. \cite{vatansever2}, it has been reported that 
the field period does not alter the compensation behavior of a mixed ferrimagnetic bulk system. In this regard, it can be concluded that the mechanism behind the  variation of the compensation behavior with respect to the stochastic dynamics 
in low dimensional systems such as magnetic bilayers may be rather different from those originated in bulk systems. 

\section{Conclusion}\label{conclusion}
We have performed Monte Carlo simulations regarding the magnetic properties of an Ising bilayer system defined on a couple of stacked honeycomb lattices where the sublattices $A$ and $B$ interact via indirect exchange coupling $J_{R}$.
In the first part of our analysis, we have investigated the equilibrium ferrimagnetic properties of the system, and we obtained $P$-, $N$-, $Q$- type magnetization profiles which have been classified according to 
Ne\'{e}l classification scheme. Compensation phenomenon suddenly disappears with decreasing strength of indirect ferrimagnetic interlayer exchange coupling. 
We have also compared the obtained results with those reported in the literature, and found that 
MC simulations qualitatively reproduce the magnetization curves obtained from EFT. In this regard, we have concluded that  EFT method exhibits the same topology as those obtained from
the MC simulation with less computational time. In the second part of our analysis, we have focused on the evolution of compensation behavior observed in the system in the presence of a time depending magnetic field.
Two different forms for the time dependence of the dynamic magnetic field has been considered as sinusoidal oscillations, and square wave form. For both cases, compensation point $T_{comp}$ and transition temperature $T_{c}$ tend to decrease with
increasing field amplitude $h_{0}$. The increasing field period $P$ also causes to the same consequence. For the fixed values of $h_{0}$ and $P$, obtained $T_{comp}$ and $T_{c}$ values for a square wave 
are clearly lower than those obtained for the sinusoidally oscillating 
magnetic fields.

Investigation of dynamical critical properties  of magnetic spin systems revealed very rich physical phenomena, and these systems promise even more interesting and novel features. For instance, whether the critical exponents of magnetization
and magnetic susceptibility exhibit any dimensional crossover as the geometry of the kinetic Ising bilayer system evolves from graphene-like structure to a graphite-like topology seems to be an interesting problem. 
However, this will be the subject of our near future work.   

\section*{Acknowledgements}
The numerical
calculations reported in this paper were performed
at TUBITAK ULAKBIM High Performance and Grid
Computing Center (TR-Grid e-Infrastructure).

\end{document}